\documentstyle[12pt,epsfig]{article}
\let\section=\subsection     \let\subsection=\subsubsection                %%
\begin{document}
\begin{center}
   {\large \bf Quantum interference in the $e^+e^-$ decays of $\rho^0$ and}\\[2mm]
   {\large \bf $\omega$-mesons produced in $\pi^-$p reactions}\\[5mm]
   Madeleine~Soyeur$^1$, Matthias~Lutz$^2$ and Bengt~Friman$^{2,3}$ \\[5mm]
   {\small \it 1.  D\'epartement d'Astrophysique, de Physique des Particules,\\
   de Physique Nucl\'eaire et de l'Instrumentation Associ\'ee\\
   Service de Physique Nucl\'eaire\\
   Commissariat \`a l'Energie Atomique/Saclay\\
   F-91191 Gif-sur-Yvette Cedex, France\\[1mm]}
   {\small \it 2. GSI, Planckstrasse 1,  D-64291 Darmstadt, Germany \\[1mm] }
   {\small \it 3.  Institut f\"ur Kernphysik, TU Darmstadt \\
   D-64289 Darmstadt, Germany \\[8mm] }

\end{center}

\begin{abstract}\noindent
   The study of the $\pi^-p \rightarrow \rho^0 n$ and
$\pi^-p \rightarrow \omega n$ amplitudes close and below the
vector meson production threshold ($1.2<\sqrt s <1.8$ GeV)
reveals a rich dynamics arising from the presence of specific
baryon resonances in this energy range. The interference pattern of
the $e^+e^-$ decays
of the $\rho^0$- and $\omega$-mesons produced in
$\pi^-$p reactions reflects directly this dynamics. We discuss
this interference pattern
in the $\pi^-p \rightarrow e^+e^- n$ reaction as function of the
 total center of mass energy $\sqrt s$. We emphasize the importance
of an experimental study of this reaction, which could be made with
the HADES detector and the available pion beam at GSI.

\end{abstract}

\section{Introduction}

The coupling of light vector mesons [$\rho$(770) and $\omega$(782)]
to low-lying baryon resonances is still to a large extent unknown.
This lack of information is a particularly important source of
uncertainties in the theoretical description of the propagation of
vector mesons in a nuclear medium, where resonance-hole states are
expected to contribute largely to the dynamics.

The $\pi^-p \rightarrow \rho^0 n$ and $\pi^-p \rightarrow \omega n$
processes have been described recently in the framework of a
relativistic coupled-channel model \cite{Lutz1}. They are
particular processes included in a broader scheme aiming at
reproducing data on pion-nucleon elastic scattering and pion-induced
production reactions invol\-ving the $\pi \Delta$, $\rho$N,
$\omega$N, K$\Lambda$, K$\Sigma$ and $\eta$N channels. The
model is restricted to s-wave scattering in the $\rho$N and
$\omega$N channels. The corresponding s- and d-wave resonances in
the $\pi$N channel are generated dynamically. The meson-baryon
coupling strengths are determined from the fit to the available
data on the channels included in the calculation.

The  $\pi^-p \rightarrow \rho^0 n$ and $\pi^-p
\rightarrow \omega n$ amplitudes are very sensitive to the pre\-sence of the
s- and d-wave pion-nucleon resonances lying below the vector meson
production threshold ($1.3<\sqrt s <1.7$ GeV). This point is
discussed and illustrated in Section 2. Data that directly reflect
these amplitudes would provide very useful constraints on the
underlying dynamics. The $\pi^-p
\rightarrow e^+e^- n$ reaction appears as a particularly relevant
process to study the $\pi^-p
\rightarrow \rho^0 n$ and $\pi^-p
\rightarrow \omega n$ amplitudes.
This reaction offers the possibility to test experimentally the
$\rho^0$ and $\omega$ strengths below threshold and the quantum
interference in the $e^+e^-$ decays of the $\rho^0$- and
$\omega$-mesons is very sensitive to the magnitudes and the
relative phase of the production amplitudes. In Section 3 we
present briefly the formalism
%of our calculation of the  $\pi^-p
%\rightarrow e^+e^- n$ reaction
and preliminary numerical results.
The perspectives of this work are discussed in Section 4.

\section{The  $\pi^-p \rightarrow \rho^0 n$ and $\pi^-p
\rightarrow \omega n$ amplitudes close to the vector
meson production threshold}

The  $\pi^-p \rightarrow \rho^0 n$ and $\pi^-p
\rightarrow \omega n$ amplitudes of Ref. \cite{Lutz1}
entering our calculation
of the $\pi^-p \rightarrow e^+e^- n$ reaction are displayed
in Fig. 1. We shall restrict our discussion to $e^+e^-$
pairs of invariant masses ranging from 0.5 to 0.8 GeV.
The exclusive measurement of the $e^+e^- n$ outgoing channel
ensures that the $e^+e^-$ pairs come from vector
meson decays (pseudoscalar mesons decay into
an $e^+e^-$ pair and an additional photon).
We recall however that only s- and d-wave pion-nucleon resonances
are at present included in the model of Ref. \cite{Lutz1}.
To be complete, the description of the  $\pi^-p \rightarrow e^+e^- n$
reaction in the energy range discussed in this work ($1.2<\sqrt s <1.8$ GeV)
should include also the effect of
%p-wave pion-nucleon resonances.
other partial waves. We will return to this question in Section 4.

The  $\pi^-p \rightarrow \rho^0 n$ and $\pi^-p
\rightarrow \omega n$ scattering amplitudes of Fig. 1 illustrate
the importance of baryon resonances in vector meson production
below threshold. These resonances induce a rich structure in both
the real and imaginary parts of the amplitudes. In particular, the
presence of the d-wave N*(1520) resonance is clearly reflected in
the J=3/2 amplitudes for $\rho^0$ and $\omega$ production. This is
an immediate consequence of the strong coupling of the N*(1520) to
both the $\rho^0 n$ and $\omega n$ channels \cite{Lutz1}. The
strong couplings imply that there is considerable vector-meson
strength in the N*-hole modes in the nuclear medium.

\newpage
\begin{figure}[t]
\begin{center}
\mbox{\epsfig{file=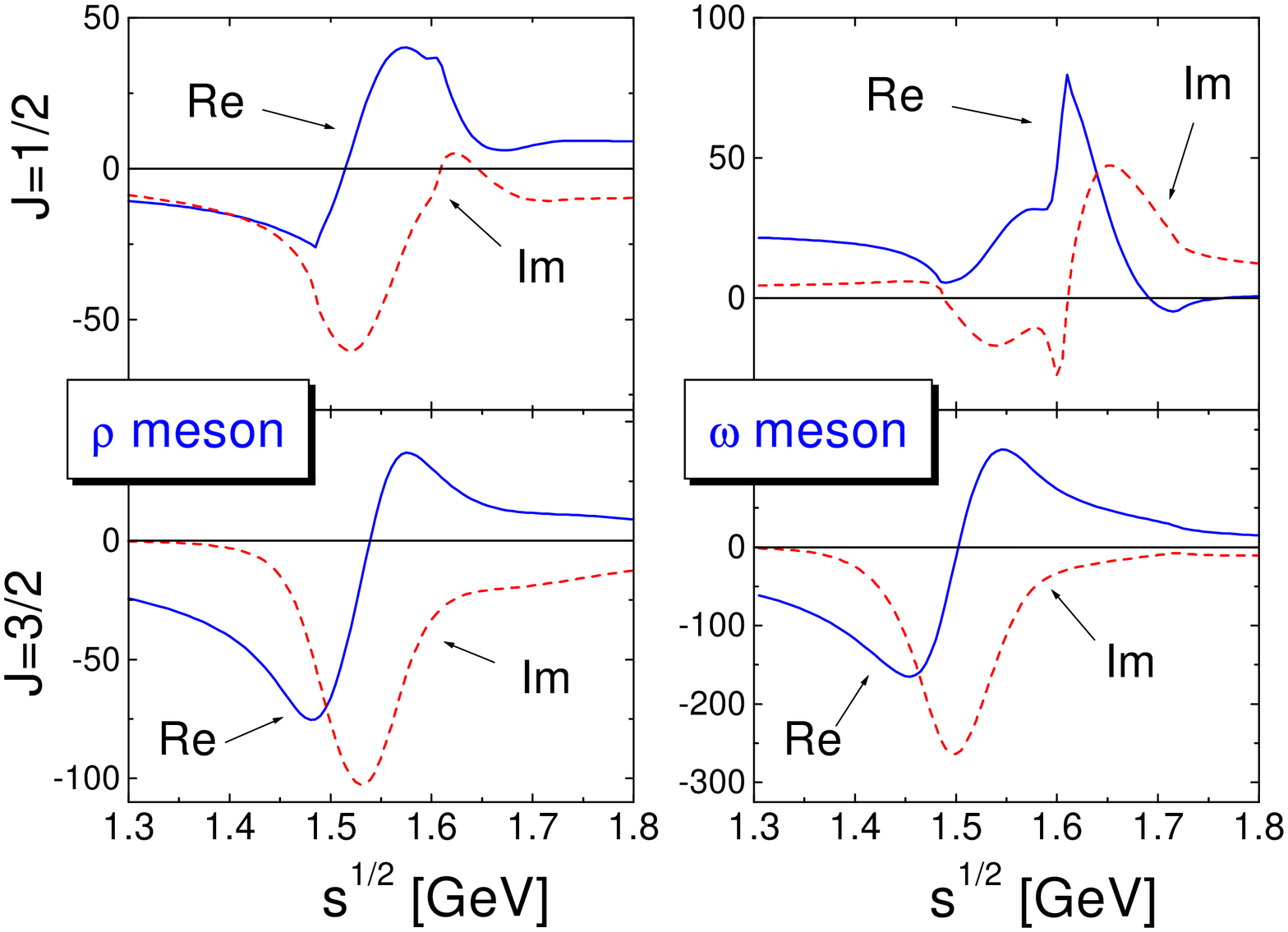, height=9.5cm}}
\end{center}
\noindent
{\begin{small}Fig.~1. Amplitudes in GeV$^{-1}$
for the  $\pi^-p \rightarrow \rho^0 n$ and $\pi^-p
\rightarrow \omega n$ processes obtained in Ref. [1].
The amplitudes are averaged over isospin and shown
for the two spin channels. \end{small}}
\end{figure}
\vskip 0.4 truecm

An experimental
test of the  N*N$\rho^0$ and N*N$\omega$
vertices through the $\pi^-p \rightarrow e^+e^- n$ reaction
below the vector meson production threshold would be a most
valuable constraint on the in-medium propagation of
$\rho^0$- and $\omega$-mesons.

\section{The  $\pi^-p \rightarrow  e^+e^- n$ reaction}

The  $\pi^-p \rightarrow \rho^0 n$ and $\pi^-p
\rightarrow \omega n$ amplitudes are simply related to the
$\pi^-p \rightarrow  e^+e^- n$ amplitudes through the Vector
Dominance assumption \cite{Sakurai,Kroll}. In this picture, the $
e^+e^-$ decay of vector mesons is described by their conversion
into time-like photons which subsequently materialize into $e^+e^-$
pairs. The magnitude of the coupling constants $f_\rho$ and
$f_\omega$, which characterize the conversion of $\rho$- and
$\omega$-mesons into photons, is determined from the measured
partial widths of $\rho^0$- and $\omega$-mesons into $e^+e^-$ pairs
\cite{PDG}. The relative phase of the $\rho$ and $\omega$
amplitudes is not determined by hadronic observables. We
determine this phase in each channel by comparing with the
photon-decay helicity amplitudes of the corresponding
resonance~\cite{PDG}, assuming Vector Meson Dominance. We use
$f_\rho$=0.036 GeV$^2$ and $f_\omega$=0.011 GeV$^2$ \cite{Friman1}.

%\newpage
\begin{figure}[h]
\begin{center}
\mbox{\epsfig{file=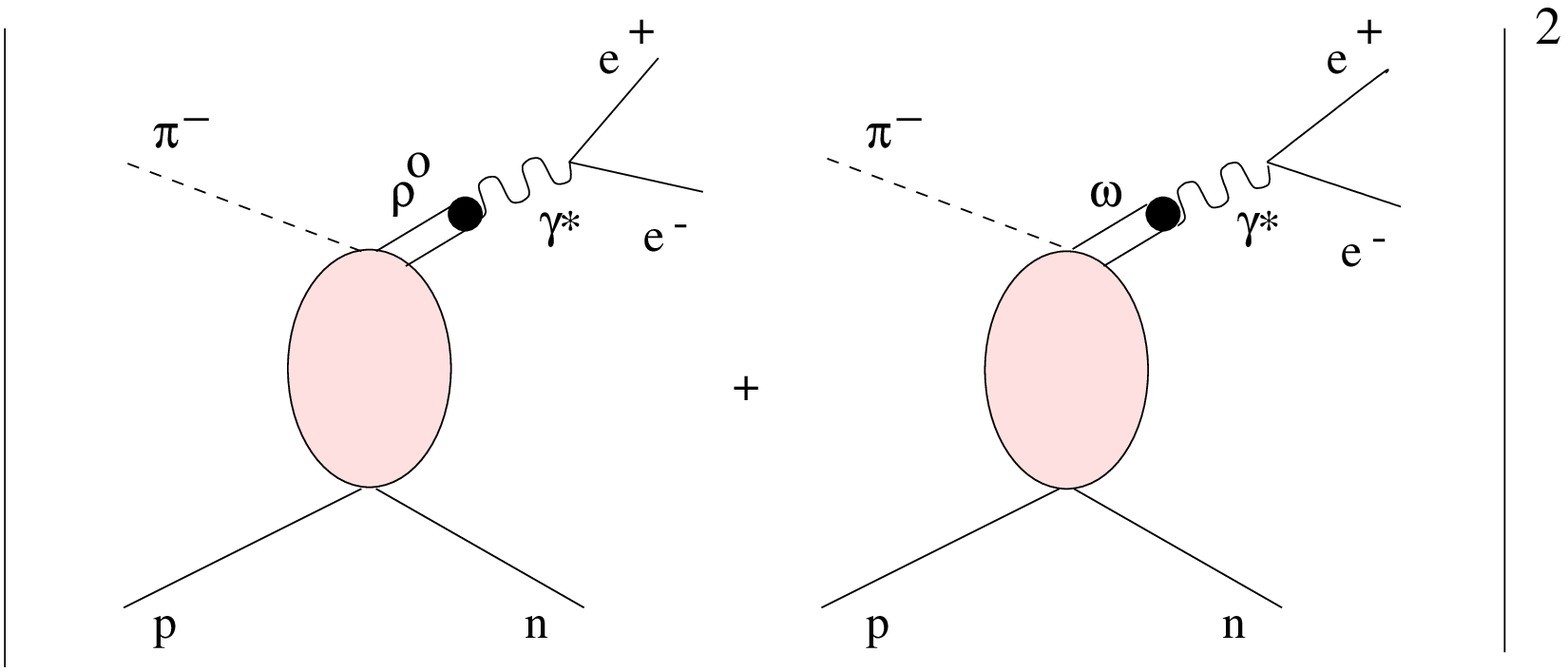, height=5cm}}
\end{center}
{\begin{small}Fig.~2. Squared amplitude for the $\pi^-p \rightarrow  e^+e^- n$
reaction with intermediate $\rho^0$- and $\omega$-mesons. \end{small}}
\end{figure}
\vskip 0.3 truecm

The squared amplitude for the $\pi^-p \rightarrow  e^+e^- n$
reaction with intermediate $\rho^0$- and $\omega$-mesons in the
Vector Dominance Model is illustrated in Fig. 2. Schema\-tically,
this quantity can be written as
\begin{eqnarray}
 \left |<n e^+e^-|\pi^-p>\right|^2
=\frac {\left |<e^+e^-|\gamma*>\right|^2} {m^4}
\mid \frac {f_{\rho}\ {\cal M}_{\pi^-p \rightarrow \rho^0 n}}
{m^2 -m_\rho^2+im_\rho\Gamma_\rho(m)} \nonumber \\
 +\, \frac {f_{\omega}\ {\cal M}_{\pi^-p \rightarrow \omega n}}
{m^2 -m_\omega^2+im_\omega\Gamma_\omega(m)}\mid\, ^2,
\end{eqnarray}
\noindent
where the first term of the right-hand side describes the
propagation of the time-like photon and its decay into an  $e^+e^-$
pair of invariant mass m and the second term contains the vector
meson production dynamics. The vector mesons are characterized by
their mass m$_V$ and energy-dependent width $\Gamma_V(m)$. The
interference of the complex ${\cal M}_{\pi^-p \rightarrow
\rho^0 n}$ and ${\cal M}_{\pi^-p \rightarrow \omega n}$ amplitudes
(Fig. 1) in the $\pi^-p \rightarrow  e^+e^- n$ cross section is
sensitive to their relative phase. The importance of measuring such a
phase in the $e^+e^-$ or $\pi^+ \pi^-$ decays of $\rho^0$- and
$\omega$-mesons has been evidenced by the contribution of such data
to the understanding of other processes, like the photoproduction
of $\rho^0$- and $\omega$-mesons in the diffractive regime ($\gamma
Be \rightarrow  e^+e^- Be$) \cite{Alvensleben} and the $ e^+e^-
\rightarrow \pi^+ \pi^-$ reaction \cite{Benakas,Connell}.

We indicate the magnitude of the $\rho^0-\omega$ interference in
the $\pi^-p \rightarrow  e^+e^- n$ reaction as function of the
total center of mass energy in Fig. 3. We have selected $e^+e^-$
pairs of invariant mass m=0.55 GeV. This figure illustrates the
role of baryon resonances with masses in the range of 1.5 to 1.6
GeV in generating strong interference effects.

\noindent
\begin{figure}[t]
\begin{center}
\mbox{\epsfig{file=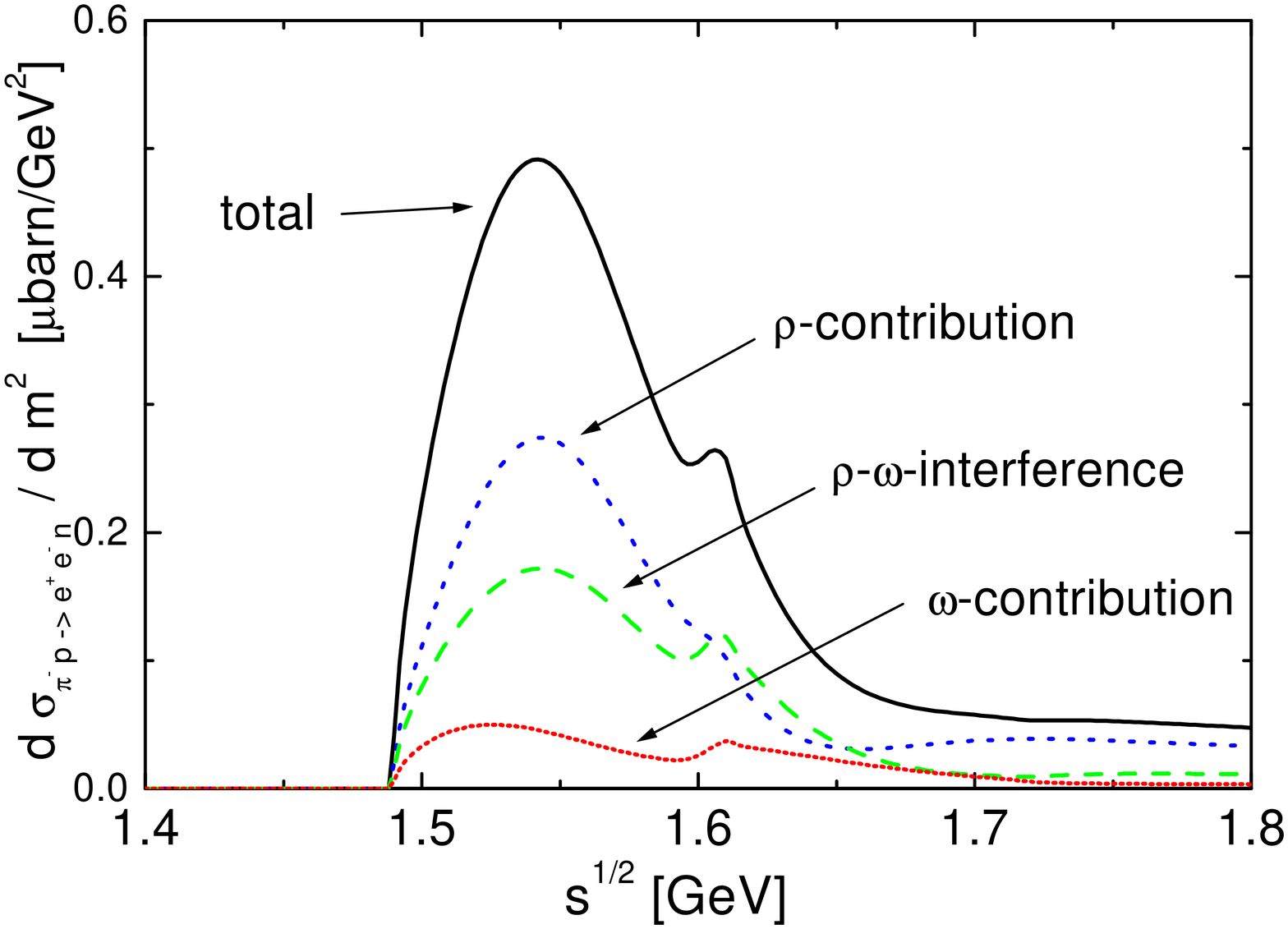, width=14.5cm}}
\end{center}
\noindent
{\begin{small}Fig.~3. Differential cross section d$\sigma$/dm$^2$
for the  $\pi^-p \rightarrow e^+e^- n$ reaction as function
of $\sqrt s$ for a fixed $ e^+e^-$ pair invariant mass m=0.55 GeV.
\end{small}}
\end{figure}

Above the vector meson threshold, the  $\rho^0-\omega$ interference
in the  $\pi^-p \rightarrow e^+e^- n$ cross section
is particularly interesting for  $ e^+e^-$ pair invariant masses
close to the $\omega$ mass. This effect is manifested in the
 invariant mass spectrum displayed in Fig. 4 ($\sqrt s$=1.8 GeV).
The model of Ref. \cite{Lutz1} for the
 ${\cal M}_{\pi^-p \rightarrow \rho^0 n}$
and ${\cal M}_{\pi^-p \rightarrow \omega n}$ amplitudes
predicts a constructive interference at this energy.
This feature appears to be a very sensitive test of the
model.
\newpage

A detailed discussion of these interference patterns
will be presented in a forthcoming publication \cite{Lutz2}.

\section{Perspectives}

The study of the $\pi^-p \rightarrow e^+e^- n$ reaction provides
a particularly stringent test of the
$\pi^-p \rightarrow \rho^0 n$ and
$\pi^-p \rightarrow \omega n$ amplitudes close and below the
vector meson production threshold ($1.2<\sqrt s <1.8$ GeV).

\noindent
\begin{figure}[t]
\begin{center}
\mbox{\epsfig{file=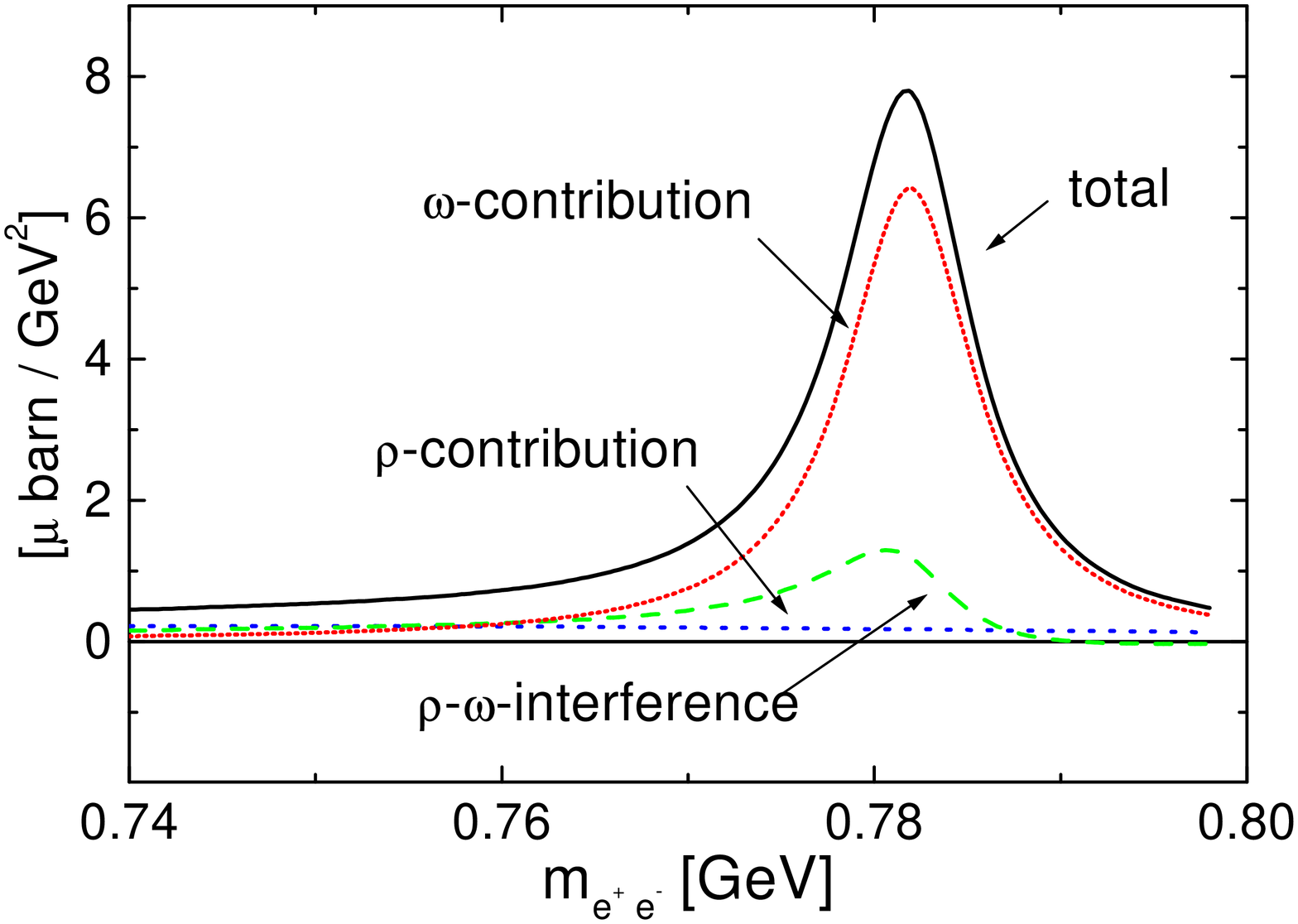, width=14.5cm}}
\end{center}
\noindent
{\begin{small}Fig.~4. Differential cross section d$\sigma$/dm$^2$
for the  $\pi^-p \rightarrow e^+e^- n$ reaction as function
of the $e^+e^-$ pair invariant mass for a fixed total center
of mass energy $\sqrt s$=1.8 GeV.\end{small}}
\end{figure}

We have computed the cross section of the  $\pi^-p \rightarrow
e^+e^- n$ reaction  using the model of Ref. \cite{Lutz1} for the
vector meson production amplitude and indicated its main features
as function of the total center of mass energy.

\newpage
A natural extension of the present work would be
to include the p-wave pion-nucleon resonances in the coupled channel
scheme of Ref. \cite{Lutz1}, thereby increasing the expected domain
of validity of the
$\pi^-p \rightarrow \rho^0 n$ and
$\pi^-p \rightarrow \omega n$ amplitudes.
Projecting the coupled-channel amplitudes on specific
s- and t-channel exchanges could be a useful step
in providing a simple interpretation of our numerical results.

We note that the study of the quantum interference of $\rho^0$- and
$\omega$-mesons produced in the  $\pi^-p \rightarrow \rho^0 n$ and
$\pi^-p \rightarrow \omega n$ reactions in other channels than the
$e^+e^-$ decay ($\pi^0 \gamma$ for example) may also be of
interest.

Data on the $\pi^-p \rightarrow e^+e^- n$ cross section in the
energy range considered in this work are at present not available.
Such measurements would provide an important test of the dynamics in a
reaction which is crucial for the understanding of the in-medium
propagation of vector mesons.

\section{Acknowledgements}

One of us (M. S.) acknowledges the generous hospitality of the
Theory Group of GSI, where much of this work was done.

%\newpage

\end{document}